\documentstyle[preprint,aps]{revtex} 

\begin{document} 

\draft 

\preprint{UTPT-95-27} 

\title{Cosmological Models in the Nonsymmetric Gravitational Theory}

\author{J. W. Moffat} 

\address{Department of Physics, University of Toronto,
Toronto, Ontario, Canada M5S 1A7} 

\date{\today}

\maketitle 

\begin{abstract}%
Cosmological consequences of the nonsymmetric gravitational theory (NGT) are
studied. The structure of the NGT field equations is analyzed for an
inhomogeneous and anisotropic universe, based on the spherically symmetric
field equations. It is assumed that the matter density and pressure are  purely time dependent, and
it is shown that the field equations allow open, flat and closed universes. The field equations are
expanded about the Friedmann-Robertson-Walker (FRW) model for a small antisymmetric field.
The observations are most in accord with the approximately spatially flat universe, which has an
FRW metric to lowest order in the skew field. The recent analysis of galaxy synchroton radiation
polarization data, which indicates a residual polarization after Faraday polarization is subtracted,
can be explained by the birefringence of electromagnetic waves produced by the NGT spacetime.

\end{abstract} 

\pacs{ } 

\narrowtext 

\section{Introduction}

A new version of the nonsymmetric gravitational theory (NGT) has been
published
\cite{Moffat1,Moffat2,Moffat3,Moffat4,LegareMoffat1,LegareMoffat2,Clayton1,Clayton2,Clayton3,Damour},
which possesses a linear approximation
free of ghost poles and tachyons, with well-behaved asymptotic conditions and a stable
expansion about a general relativistic (GR) background to first order in the skew field
$g_{[\mu\nu]}$.

The standard model scenario in big bang cosmology assumes that at any given time
the universe is homogeneous and isotropic when averaged over a sufficiently
large scale. The equations of the standard model can be expressed as
\begin{equation}
H^2+\frac{k}{R^2}=\Omega_M H^2+\Omega_\Lambda H^2,
\end{equation}
where $H=\dot{R}/R$ is the Hubble parameter, $\Omega_M$ is the
density parameter for matter and $\Omega_\Lambda$ is the density parameter
contributed by the cosmological constant.  This model gives a good account of present
experimental data\cite{Peebles}, although there are indications that the latest data for the age of
the universe may require a non-zero cosmological constant\cite{Moffat6}.

In the following, we shall derive a cosmological model in which it is assumed that the skew
$g_{[\mu\nu]}$ contributions are small in the present universe, so that we can expand the NGT
field equations about the homogeneous and isotropic Friedmann-Robertson-Walker (FRW)
model and obtain an approximate dynamical equation for the cosmological scale factor $R(t)$.
We base the cosmological solutions on an inhomogeneous spherically symmetric
$g_{\mu\nu}$, and assume for simplicity that the matter density $\rho_M$ and the pressure $p$
are uniform and only depend on the time $t$. For the case of a spatially flat universe, the metric
takes the simple form of an FRW universe to lowest order in $g_{[\mu\nu]}$  and the
Friedmann-type equation for $R(t)$ contains corrections due to a non-zero $g_{[\mu\nu]}$
and its derivatives.

Savaria\cite{Savaria} obtained an exact plane
symmetric anisotropic solution both for the vacuum field equations and for the field
equations in the presence of pressureless matter.

\section{Structure of the NGT Field Equations}

We shall decompose the nonsymmetric $g_{\mu\nu}$ and
$\Gamma^\lambda_{\mu\nu}$ as
\begin{equation}
g_{(\mu\nu)}={1\over 2}(g_{\mu\nu}+g_{\nu\mu}),\quad g_{[\mu\nu]}=
{1\over 2}(g_{\mu\nu}-g_{\nu\mu}),
\end{equation}
and
\begin{equation}
\Gamma^\lambda_{\mu\nu}=\Gamma^\lambda_{(\mu\nu)}
+\Gamma^\lambda_{[\mu\nu]}.
\end{equation}
The contravariant tensor $g^{\mu\nu}$ is defined in terms of the equation:
\begin{equation}
\label{inverse}
g^{\mu\nu}g_{\sigma\nu}=g^{\nu\mu}g_{\nu\sigma}={\delta^\mu}_\sigma.
\end{equation}

We shall use the notation: ${\bf g}^{\mu\nu}=\sqrt{-g}g^{\mu\nu}$,
$g=\hbox{Det}(g_{\mu\nu})$, and $R_{\mu\nu}(W)$ is the NGT contracted curvature tensor: 
\begin{equation}
R_{\mu\nu}(W)=W^\beta_{\mu\nu,\beta} 
- {1\over 2}(W^\beta_{\mu\beta,\nu}+W^\beta_{\nu\beta,\mu}) - 
W^\beta_{\alpha\nu}W^\alpha_{\mu\beta} +
W^\beta_{\alpha\beta}W^\alpha_{\mu\nu},
\end{equation}
defined in terms of the unconstrained nonsymmetric connection:
\begin{equation}
\label{Wequation}
W^\lambda_{\mu\nu}=\Gamma^\lambda_{\mu\nu}-{2\over 3}{\delta^\lambda}_\mu
W_\nu,
\end{equation}
where 
\[
W_\mu={1\over 2}(W^\lambda_{\mu\lambda}-W^\lambda_{\lambda\mu}).
\]
Eq.(\ref{Wequation}) leads to the result: 
\[
\Gamma_\mu=\Gamma^\lambda_{[\mu\lambda]}=0.
\]
The contracted tensor $R_{\mu\nu}(W)$ can be written as
\[
R_{\mu\nu}(W)=R_{\mu\nu}(\Gamma)+\frac{2}{3}W_{[\mu,\nu]},
\]
where
\[
R_{\mu\nu}(\Gamma ) = \Gamma^\beta_{\mu\nu,\beta} -{1\over 2} 
\left(\Gamma^\beta_{(\mu\beta),\nu} + \Gamma^\beta_{(\nu\beta),\mu}\right) - 
\Gamma^\beta_{\alpha\nu} \Gamma^\alpha_{\mu\beta} + 
\Gamma^\beta_{(\alpha\beta)}\Gamma^\alpha_{\mu\nu}.
\]

The field equations in the new version of NGT take the form\cite{Moffat1}:
\begin{mathletters}
\begin{eqnarray}
\label{Gequation}
G_{\mu\nu} (W)+\Lambda g_{\mu\nu}+S_{\mu\nu}
&=&8\pi (T_{\mu\nu}+K_{\mu\nu}),\\
\label{divg}
{{\bf g}^{[\mu\nu]}}_{,\nu}&=&-\frac{1}{2}{\bf g}^{(\mu\alpha)}W_\alpha,\\
\label{gskewconstraint}
{\bf g}^{[\mu\nu]}u_\nu=0,\\
{{\bf g}^{\mu\nu}}_{,\sigma}+{\bf g}^{\rho\nu}W^\mu_{\rho\sigma}
+{\bf g}^{\mu\rho}
W^\nu_{\sigma\rho}-{\bf g}^{\mu\nu}W^\rho_{\sigma\rho}
+{2\over 3}\delta^\nu_\sigma{\bf g}^{\mu\rho}W^\beta_{[\rho\beta]}\nonumber \\
+{1\over 6}({\bf g}^{(\mu\beta)}W_\beta\delta^\nu_\sigma
-{\bf g}^{(\nu\beta)}W_\beta\delta^\mu_\sigma)&=&0.
\end{eqnarray}
\end{mathletters}
Here, we have
\begin{mathletters}
\begin{eqnarray}
G_{\mu\nu}&=&R_{\mu\nu} - {1\over 2} g_{\mu\nu} R,\\
S_{\mu\nu}&=&-\frac{1}{6}(W_\mu W_\nu-\frac{1}{2}g_{\mu\nu}g^{\alpha\beta}W_\alpha
W_\beta),
\end{eqnarray}
\end{mathletters}
$T_{\mu\nu}$ denotes the nonsymmetric source tensor and $\Lambda$ is the cosmological
constant. Moreover, using 
Eq.(\ref{gskewconstraint}) we have
\begin{equation}
\label{Kequation}
K_{\mu\nu}=-\frac{1}{8\pi}u_{[\mu}\phi_{\nu]},
\end{equation}
where $\phi_{\mu}$ are four Lagrange multiplier fields\cite{Moffat1},
$u_\mu=dx_\mu/d\tau$ is the four-velocity vector and $\tau$ denotes the proper time along an
observer's world line in spacetime. We impose the condition:
\begin{equation}
g^{(\mu\nu)}u_\mu u_\nu=1.
\end{equation}

We can choose the vector $u_\mu$ to be $u_\mu=(0,0,0,1/\sqrt{g^{00}})$, so that
(\ref{gskewconstraint}) corresponds to the three constraint equations:
\begin{equation}
\label{constraints2}
{\bf g}^{[i0]}=0.
\end{equation}

The generalized Bianchi identities:
\begin{equation}
[{\bf g}^{\alpha\nu}G_{\rho\nu}(\Gamma)+{\bf g}^{\nu\alpha}
G_{\nu\rho}(\Gamma)]_{,\alpha}+{g^{\mu\nu}}_{,\rho}{\bf G}_{\mu\nu}=0,
\end{equation}
give rise to the matter response equations\cite{Moffat4}:
\begin{equation}
g_{\mu\rho}{{\bf T}^{\mu\nu}}_{,\nu}+g_{\rho\mu}{{\bf T}^{\nu\mu}}_{,\nu}
+(g_{\mu\rho,\nu}+g_{\rho\nu,\mu}-g_{\mu\nu,\rho}){\bf T}^{\mu\nu}=0.
\end{equation}

After eliminating the Lagrange multiplier $\phi_\mu$ from the field equations (\ref{Gequation}),
we get
\begin{mathletters}
\begin{eqnarray}
G_{(\mu\nu)} (W)+\Lambda g_{(\mu\nu)}+S_{(\mu\nu)}&=&8\pi T_{(\mu\nu)},\\
\label{skewG}
\epsilon^{\mu\nu\alpha\beta}u_\alpha(G_{[\mu\nu]}+\Lambda g_{[\mu\nu]}+S_{[\mu\nu]})
&=&8\pi\epsilon^{\mu\nu\alpha\beta}u_\alpha T_{[\mu\nu]},
\end{eqnarray}
\end{mathletters}
where $\epsilon^{\mu\nu\alpha\beta}$ is the Levi-Civita symbol.

For the case of a spherically symmetric field, the canonical form of $g_{\mu\nu}$,
in NGT, is given by
\[
g_{\mu\nu}=\left(\matrix{-\alpha&0&0&w\cr 
0&-\beta&f\hbox{sin}\theta&0\cr 0&-f\hbox{sin}\theta&
-\beta\hbox{sin}^2
\theta&0\cr-w&0&0&\gamma\cr}\right),
\]
where $\alpha,\beta, \gamma$ and $w$ are functions of $r$ and $t$.
From the constraint equations (\ref{constraints2}), we have $w=0$ and only the $g_{[23]}$
component of $g_{[\mu\nu]}$ is different from zero.
We have
\[
\sqrt{-g}=\hbox{sin}\theta[\alpha\gamma(\beta^2+f^2)]^{1/2}.
\]

If we adopt the approximation scheme leading to the geodesic equation
for falling test particles, then we can use a comoving coordinate
system with the velocity components\cite{LegareMoffat2}: 
\[
u^0=1,\quad u^r=u^\theta=u^\phi=0,
\]
and the time dependent metric in normal Gaussian form:
\begin{equation}
ds^2=dt^2-\alpha(r,t)dr^2-\beta(r,t)(d\theta^2+\hbox{sin}^2\theta
d\phi^2).
\end{equation}

The field equations for the spherically symmetric system take the
form:
\begin{eqnarray}
\label{SSfieldequations}
G_{\mu\nu}(\Gamma)+\Lambda
g_{\mu\nu}&=&8\pi GT_{\mu\nu},\\
{{\bf g}^{[\mu\nu]}}_{,\nu}&=&0,
\end{eqnarray}

We have for the spherically symmetric conservation laws:
\begin{equation}
\label{SSconserve}
\hbox{Re}[(T_{\rho-\nu-}{\bf g}^{\sigma+\nu-})_{;\sigma}]=0,
\end{equation}
where we have used the Einstein + and - notation for covariant differentiation
with respect to the $\Gamma^\lambda_{\mu\nu}$ connection\cite{Einstein}.
We assume the energy-momentum tensor takes the form:
\begin{equation}
T^{\mu\nu}=(\rho+p)u^\mu u^\nu-pg^{\mu\nu}+N^{[\mu\nu]},
\end{equation}
where $N^{[\mu\nu]}$ is a direct source for the antisymmetric fields.

In order to simplify the field equations, we must make several 
approximations. We shall assume that the direct coupling term in the source
tensor, $N^{[\mu\nu]}$, is small and can be neglected and we also set $\Lambda=0$.

To facilitate a comparison of the NGT cosmological models with the standard FRW model, we
shall adopt the approximation:
\begin{equation}
\label{approx1}
\beta(r,t) \gg f(r,t).
\end{equation}
We obtain the field equations:
\begin{mathletters}
\begin{eqnarray}
\label{eq:simpleeqs_11}
-{1\over \alpha}\biggl[{\beta'' \over \beta}-{\beta^{' 2}\over 2\beta^2}
-{\alpha^\prime\beta^\prime\over
2\alpha\beta}\biggr]+{\ddot{\alpha}\over 2\alpha}
-{\dot\alpha^2\over {4\alpha^2}}+{\dot\alpha\dot\beta\over 2\alpha\beta}
+W&=&4\pi G(\rho_M-p), \\
\label{eq:simpleeqs_22}
{1\over \beta}-{1\over \alpha}\biggl({\beta'' \over 2\beta}
-{\alpha^\prime\beta^\prime\over 4\alpha\beta}\biggr)
+{{\ddot{\beta}}\over 2\beta}+{{\dot{\alpha}}{\dot{\beta}}\over
4\alpha\beta}+ X
&=&4\pi G(\rho_M-p), \\
\label{eq:simpleeqs_00}
-{{\ddot{\alpha}}\over 2\alpha}-{{\ddot{\beta}}\over
\beta}+{{\dot{\alpha}}^2\over
4\alpha^2}+{{\dot{\beta}}^2\over 2\beta^2}+ Y&=&4\pi G(\rho_M+3p), \\
\label{eq:simpleeqs_(01)}
-{{\dot{\beta}}^\prime\over \beta}+{\beta^\prime{\dot{\beta}}\over
2\beta^2}+{{\dot{\alpha}}\beta^\prime\over 2\alpha\beta}+ Z&=&0,\\
\label{eq:simpleeqs_[23]}
\frac{\dot f\dot\beta}{2\beta f}-\frac{\ddot f}{2f}+\frac{\beta'^2}{2\alpha\beta^2}
-\frac{{\dot\beta^2}}{2\beta^2}+\frac{\alpha'\beta'}{2\alpha^2\beta}
-\frac{\beta^{\prime\prime}}{\alpha\beta}+\frac{\ddot\beta}{\beta}
-\frac{f'\beta'}{2\alpha\beta f}
\nonumber\\
-\frac{\dot\alpha\dot f}{4\alpha f}+\frac{\dot\alpha\dot\beta}{2\alpha\beta}
+\frac{f^{\prime\prime}}{2\alpha f}-\frac{\alpha' f'}{4\alpha^2 f}&=
&4\pi G(\rho_M-p).
\end{eqnarray}
\end{mathletters}
Here, we have defined $\dot{\alpha}=d\alpha/dt$, $\alpha^\prime=d\alpha/dr$ and
\begin{mathletters}  
\begin{eqnarray}
\label{Wexpression}
W(r,t)&=&-\frac{\alpha'\beta'f^2}{2\alpha^2\beta^3}
+\frac{\beta^{\prime\prime}f^2}{\alpha\beta^3}
\nonumber\\
& &\mbox{}
-\frac{\dot\alpha\dot\beta f^2}{2\alpha\beta^3}-\frac{5\beta'^2f^2}
{2\alpha\beta^4}+\frac{\dot\alpha f\dot f}{2\alpha\beta^2}
+\frac{\alpha'ff'}{2\alpha^2\beta^2}-\frac{ff^{\prime\prime}}{\alpha\beta^2}
+\frac{4ff'\beta'}{\alpha\beta^3}-\frac{3f'^2}{2\alpha\beta^2},\\
X(r,t)&=&-\frac{\dot\alpha\dot\beta f^2}{2\alpha\beta^3}-\frac{\alpha'\beta'
f^2}{2\alpha^2\beta^3}+\frac{\beta^{\prime\prime}f^2}{\alpha\beta^3}
-\frac{\ddot\beta f^2}
{\beta^3}+\frac{\dot\alpha f\dot f}{2\alpha\beta^2}
+\frac{\dot\beta^2f^2}{\beta^4}
-\frac{ff^{\prime\prime}}{\beta^2\alpha}-\frac{\beta'^2f^2}{\alpha\beta^4}
\nonumber\\
& & \mbox{}
+\frac{\alpha'ff'}{2\alpha^2\beta^2}+\frac{\dot f^2}{2\beta^2}
+\frac{f\ddot f}{\beta^2}-\frac{f'^2}{2\alpha\beta^2}
-\frac{3f\dot f\dot\beta}{2\beta^3}+\frac{3ff'\beta'}{2\alpha\beta^3},\\
Y(r,t)&=&\frac{\ddot\beta f^2}{\beta^3}-\frac{5\dot\beta^2f^2}{2\beta^4}
-\frac{3\dot f^2}{2\beta^2}+\frac{4\dot\beta f\dot f}{\beta^3}
-\frac{f\ddot f}{\beta^2},\\
Z(r,t)&=&\frac{\dot\beta'f^2}{\beta^3}-\frac{5\dot\beta\beta' f^2}{2\beta^4}
-\frac{\dot\alpha\beta' f^2}{2\alpha\beta^3}+\frac{2\dot\beta ff'}{\beta^3}
-\frac{f\dot f'}{\beta^2}-\frac{3f'\dot f}{2\beta^2}\nonumber\\
& & \mbox{}
+\frac{\dot\alpha ff'}{2\alpha\beta^2}+\frac{2\beta'f\dot f}{\beta^3}.
\end{eqnarray}
\end{mathletters}

\section{Analysis of Cosmological Models}

An infinitesimal transformation of coordinates is described by
\begin{equation}
x^{\prime\mu}=x^\mu+\xi^\mu.
\end{equation}
The coordinate transformation law for $g_{\mu\nu}$ is given by
\begin{equation}
g^\prime_{\mu\nu}(x^\prime)=g_{\alpha\beta}(x)\frac{\partial x^\alpha}{\partial
x^{\prime\mu}}\frac{\partial x^\beta}{\partial x^{\prime\nu}},
\end{equation}
and the condition of homogeneity is
\begin{equation}
g^\prime_{\mu\nu}(x)=g_{\mu\nu}(x).
\end{equation}
We now obtain the Killing equation:
\begin{equation}
\label{Killing}
g_{\mu\sigma}{\xi^\sigma}_{,\nu}+g_{\sigma\nu}{\xi^\sigma}_{,\mu}
+g_{\mu\nu,\sigma}\xi^\sigma=0.
\end{equation}
The symmetric and skew parts of (\ref{Killing}) are given by:
\begin{equation}
g_{(\mu\sigma)}{\xi^\sigma}_{,\nu}+g_{(\sigma\nu)}{\xi^\sigma}_{,\mu}
+g_{(\mu\nu),\sigma}\xi^\sigma=0,
\end{equation}
and
\begin{equation}
\label{skewkilling}
g_{[\mu\sigma]}{\xi^\sigma}_{,\nu}+g_{[\sigma\nu]}{\xi^\sigma}_{,\mu}
+g_{[\mu\nu],\sigma}\xi^\sigma=0.
\end{equation}

Eq.(\ref{skewkilling}) gives
\begin{equation}
f{\xi^3}_{,1}=0.
\end{equation}
The Killing vectors take the form:
\begin{mathletters}
\begin{eqnarray}
\xi^1&=&q(r)[(a_1\sin\phi+a_2\cos\phi)\sin\theta+a_3\cos\theta],\\
\xi^2&=&\frac{q(r)}{r}[(a_1\sin\phi+a_2\cos\phi)\cos\theta
-a_3\sin\theta],\\
\xi^3&=&\frac{q(r)}{r\sin\theta}(a_1\cos\phi-a_2\sin\phi),\\
\xi^0&=&0,
\end{eqnarray}
\end{mathletters}
where $q(r)$ is a function that can be determined from the symmetric Killing equation. From the
form of the Killing vectors, we see that the condition of homogeneity in NGT requires that
\begin{equation}
f(r,t)=0.
\end{equation}
It follows that all strictly homogeneous and isotropic solutions in NGT cosmology reduce to
the FRW solutions of GR.

We shall simplify our model even further and assume that the
density $\rho_M$ and the pressure $p$ are independent of position. It is assumed
that a solution can be found by a separation of variables:
\begin{equation}
\label{separationeq}
\alpha(r,t)=h(r)R^2(t),\quad \beta(r,t)=r^2S^2(t).
\end{equation}
From Eq.(\ref{eq:simpleeqs_(01)}), we get
\begin{equation}
\label{RSequation}
\frac{{\dot R}}{R}-\frac{{\dot S}}{S}=\frac{1}{2}Z(r,t)r.
\end{equation}
If we assume that $Z(r,t)\approx 0$, then from (\ref{RSequation}) we find that
\[
R(t)\approx S(t).
\]
From the conservation law (\ref{SSconserve}), we obtain within our approximation
scheme:
\[
\dot{p}=R^{-3}(t)\frac{\partial}{\partial t}[R^3(t)(\rho+p)].
\]

Let us expand the metric $g_{(\mu\nu)}$ as
\begin{equation}
g_{(\mu\nu)}(r,t)=g^{HI}_{(\mu\nu)}(r,t)+\delta g_{(\mu\nu)}(r,t),
\end{equation}
where $g^{HI}_{(\mu\nu)}$ denotes the homogeneous and isotropic solution of
$g_{(\mu\nu)}$, and $\delta g_{(\mu\nu)}$ are small quantities which
break the maximally symmetric solution, $g^{\rm HI}_{(\mu\nu)}$, 
to lowest order in $g_{[\mu\nu]}$. Within this scenario, the metric line-element takes the FRW
form:
\begin{equation}
\label{FRWmetric}
ds^2=dt^2-R^2(t)\biggl[h(r)dr^2+r^2(d\theta^2+\sin^2\theta d\phi^2)\biggr].
\end{equation}
Thus, we assume that the universe is approximately isotropic and homogeneous. 
We have three distinct solutions depending on whether $h(r) <1, =1, >1$ and they
correspond to hyperbolic (open), parabolic (flat), and elliptic (closed) universes,
respectively.

Eqs.(\ref{eq:simpleeqs_11}) and (\ref{eq:simpleeqs_00}) can now be written:
\begin{eqnarray}
\label{eqn1}
2b(r)+{\ddot{R}}(t)R(t)+2{\dot{R}}^2(t)+R^2(t)W(r,t)&=&4\pi G
R^2(t)(\rho_M(t)-p(t)),\\ 
\label{eqn2}
-{\ddot R}(t)R(t)+\frac{1}{3}R^2(t)Y(t)&=&\frac{4\pi G}{3}R^2(t)
(\rho_M(t)+3p(t)),
\end{eqnarray}
where
\begin{equation}
b(r)={h^\prime(r)\over 2r h^2(r)}.
\end{equation}
Eliminating ${\ddot R}$ by adding (\ref{eqn1}) and 
(\ref{eqn2}), we get
\begin{equation}
\label{Rvelocityeq}
{\dot{R}}^2(t)=-b(r)+{8\pi G\over 3}[\rho_M(t)-\frac{3}{8\pi G}Q(r,t)]R^2(t),
\end{equation}
where
\[
Q(r,t)=\frac{1}{2}[W(r,t)+\frac{1}{3}Y(t)].
\]

We can write Eq. (\ref{Rvelocityeq}) as
\begin{equation}
H^2(t)+\frac{b(r)}{R^2(t)}=\frac{8\pi G}{3}[\rho_M(t)-\frac{3}{8\pi G}Q(r,t)].
\end{equation}
This equation can in turn
be written as
\[
H^2(t)+\frac{b(r)}{R^2(t)}=\Omega(r,t)H^2(t),
\]
where 
\[
\Omega(r,t)=\Omega_M(t)+\Omega_S(r,t),
\]
and
\[
\Omega_M(t)=\frac{8\pi G\rho_M(t)}{3H^2(t)},\quad \Omega_S(r,t)=-\frac{Q(r,t)}{H^2(t)}.
\]
Here, $\Omega_M$ denotes the familiar density parameter for matter, while
$\Omega_S$ denotes the density parameter associated with the skew field
contributions. 

If $b(r)=0$, then  we get $\Omega(r,t)=1$ and
\begin{equation}
H^2(t)=\frac{8\pi G}{3}[\rho_M(t)-\frac{3}{8\pi G}Q(t)].
\end{equation}

The line element now takes the approximate form of a flat, homogeneous and isotropic FRW
universe:
\begin{equation}
ds^2=dt^2-R^2(t)[dr^2+r^2(d\theta^2+\sin^2\theta d\phi^2)].
\end{equation}

The field equations, Eq.(21), will determine $f(r,t)=g_{[23]}/\sin\theta$,
which in turn will determine the function $b(r)$, given the matter density $\rho_M$.
The present observations favour an approximately flat universe with $\Omega_0\approx 1$.

A recent publication indicates that electromagnetic waves emitted in the form of synchrotron
radiation from galaxies propagating over cosmological distances may suffer a
residual polarization effect beyond the Faraday effect due to cosmic electromagnetic
fields\cite{Nodland}. If upheld, these observations would have a possible explanation
in the existence of an anisotropic effect due to the presence of the $g^{[23]}$ field component in
the NGT-Maxwell action\cite {MoffatMann1,Will,MoffatMann2,Haugan1,Haugan2}:
\begin{equation}
I_{\rm em}=-\frac{1}{16\pi}\int d^4x\sqrt{-g}g^{\mu\alpha}g^{\nu\beta}F_{\mu\nu}
F_{\alpha\beta},
\end{equation}
where $F_{\mu\nu}$ is the electromagnetic field strength.

The speed of light is not the same in every local inertial frame, breaking the
strong equivalence principle in GR.
Thus, the NGT spacetime acts like a crystal that produces birefringence as electromagnetic
radiation passes through it. The residual rotation angle predicted by NGT is
\begin{equation}
\beta_S\approx\frac{1}{2}\Lambda_S(r,t)\cos^2(\gamma),
\end{equation}
where $\gamma$ is the angle
between the propagating wave vector and the anisotropic axis fit to the data. 
A derivation and some consequences of the predicted polarization of electromagnetic
waves due to the cosmological NGT spacetime will be discussed in a separate
article\cite{MoffatSavaria}.

\acknowledgements

This work was supported by the Natural Sciences and Engineering Research
Council of Canada. I thank P. Savaria for helpful discussions.

\end{document}